\documentclass[%
 reprint,
 amsmath,amssymb,
 aps,
]{revtex4-1}
\usepackage[utf8]{inputenc}
\usepackage{amsmath}
\usepackage{amsfonts}
\usepackage{amssymb}
\usepackage{graphicx}
\usepackage{soul}
\usepackage{parskip}
\usepackage{natbib}
\usepackage{bbm}
\usepackage{algorithm}
\usepackage{bm}
\usepackage[noend]{algpseudocode}
\usepackage{booktabs}

\begin{document}

\title{Simulating the Spread of Epidemics in China on the Multi-layer Transportation Network: Beyond the Coronavirus in Wuhan}

\author{Tianyi Li}
\thanks{tianyil@mit.edu}
\affiliation{System Dynamics Group, Sloan School of Management, Massachusetts Institute of Technology}

\date{\today}

\begin{abstract}
Based on the SEIR model and the modeling of urban transportation networks, a general-purpose simulator for the spread of epidemics in Chinese cities is built. The Chinese public transportation system between over 340 prefectural-level cities is modeled as a \textit{multi-layer} \textit{bi-partite} network, with layers representing different means of transportation (airlines, railways, sail routes and buses), and nodes divided into two categories (central cities, peripheral cities). At each city, an open-system SEIR model tracks the local spread of the disease, with population in- and out-flow exchanging with the overlying transportation network. The model accounts for (1) different transmissivities of the epidemic on different transportation media, (2) the transit of inbound flow at cities, (3) cross-infection on public transportation vehicles due to path overlap, and the realistic considerations that (4) the infected population are not entering public transportation and (5) the recovered population are not subject to repeated infections. The model could be used to simulate the city-level spread in China (and potentially other countries) of an arbitrary epidemic, characterized by its basic reproduction number $R_0$, incubation period $D_E$, infection period $D_I$ and zoonotic force $z$, originated from any Chinese prefectural-level city(s), during the period before effective government interventions are implemented. Flowmaps are input into the system to trigger inter-city dynamics, assuming different flow strength, determined from empirical observation, within/between the bi-partite divisions of nodes. The model is used to simulate the 2019 Coronavirus epidemic in Wuhan; it shows that the framework is robust and reliable, and simulated results match public city-level datasets to an extraordinary extent. This simulator provides great resolution for epidemics studies based on transportation networks, and may be useful for future policy decision-making.
\end{abstract}

\maketitle


The 2019-nCov epidemic, originated from Wuhan, China \citep{Waet2020,Zet2020,Zhet2020,GM2020} has incurred heavy casualties and tremendous economic loss. With the first case confirmed as early as Dec. 8th, 2019 \citep{Let2020} and effective measures implemented in a national scale in China around 50 days later, the new Coronavirus has claimed over 70,000 cases and more than 2700 deaths in China (as on Feb. 26th). The epidemic took place right before the Chinese New Year, and the massive population flow across the entire country aggravated its vicious spread. Not long after its burst in the Hubei Province in early January, the disease started to propagate worldwide, invading almost all countries in East Asia and soon reached Europe, America and Australia \citep{Hoet2020,Phet2020}, due to the fact that Wuhan is a top-10 Chinese metropolitan and technically the single transportation hub in the central region of China, with a population as large as 11 millions. Facing the severe threat of the epidemic, since late January, the Chinese government has taken strong measures to quench the disease, mobilizing and coordinating available forces in a unprecedented scale, and thanks to the collective efforts of Chinese citizens, the epidemic has been largely put under control in most Chinese provinces by mid-February. Nevertheless, although the situation in China is picking up, recent reports indicate that the virus has begun to trigger vicious dynamics in other countries, and the epidemic will inevitably leave profound effects on the global economy. 
 
Right after the burst of 2019-nCov, studies set out to model the epidemic and use simulation results to nowcast and forecast its intensity \citep[e.g.,][]{Liuet2020,MM2020,Tet2020}. The 2019 Coronavirus is compared with the SARS in 2003, and people discovered that this time the virus is more contagious but less fatal. Arguably, the incubation period of 2019-nCov is believed to be longer than SARS, and the early symptoms are less salient \citep{Het2020}; to a large extent, it is for these two reasons that this epidemic was not paid enough attention to in its early period, and an exponential growth after the sufficient breeding soon took off during the Chinese New Year. \citet{Wet2020} is the first formally published simulation model for 2019-nCov. Because this disease is originated only in Wuhan and is carried to other cities, \citet{Wet2020} assembles an SEIR model for Wuhan with population in- and outflows, treating infections outside Wuhan as imported cases. The model is calibrated with the time series data of the confirmed cases in a list of global cities, and a posterior estimate of the basic reproduction number $R_0=2.68$ is carried out. Simulation models similar to the flavor of \citet{Wet2020} are of great importance for policy decision-making, whose findings may significantly help guide the rescue and response for the epidemic.

Nevertheless, a few problems arise with the model in \citet{Wet2020}, hampering it to reach sufficient resolution. One big concern is that the model does not differentiate means of transportation during the population flow. The public transportation system is composed of airlines, railways, sail routes and buses (highways). It is very important to note that especially in China where public transportation is exploited to a great extent, different means of transportation leads to different contact rates, and thus different transmissivities of the disease during travel. The spread of epidemic is substantially easier on trains or buses than on airplanes, while for (private) car travels, cross-infection is essentially negligible since the transportation is end-to-end. The differentiation of means of transportation is a must for high-resolution epidemic models regarding inter-city population flow since people spend non-trivial time on the route and may engage in various activities. Moreover, there exists cross-infection during travel due to path overlaps, which could be accounted for if means of transportation are not separated. Since railway, sails and bus travels are often not end-to-end, the spread of disease will likely take place along the way, among people taking the same vehicle yet having different destinations. Again, such an infection scenario is not negligible in China where the transportation system is crowded. Note that similar cross-infection concerns may arise in transportation service locations such as train stations and airports; yet concerns in these places are not as severe as during the travel where people are more densely located and have little choice of isolating from their neighbors who have unknown points of departure and destinations. 

Second, the model initiates an SEIR compartment model only in Wuhan but not in other cities. This only holds true if the disease is not sufficiently contagious, or after the quarantine procedures are successfully implemented in all other cities, and this model could not capture the local evolution dynamics of the imposed diseases, which is non-trivial at least for the other cities in Hubei Province. Ideally,  the same SEIR dynamics could be initiated in every city, contributing to the spread of the disease in a nationwide landscape. Third, the model neglects an important aspect of the population flow in China, the non-trivial occurrence of transit events in public transportation. Although highly developed, the Chinese transportation system could not (even get close to) realize cheap end-to-end travels between the over 340 prefectural-level cities, as the primarily concerned level of resolution in the epidemic spread. Yet Chinese citizens are also much more likely than citizens of other countries to travel a long distance to a major city (e.g., to make a living) due to an imbalanced development across the nation, which gives rise to the non-trivial role of transfer during inter-city public transportation. Without taking care of the transit issue, the direct-import model may possess a fundamental system error. Last, the model in \citet{Wet2020} allows population out-flow of Wuhan from all four SEIR compartments. This may contain another system error as it is more appropriate to assume that the infected population are not traveling but instead stay local. Also, since the model did not trace the source of the population inflow, it has to put the recovered population into the susceptible compartment and essentially assumes repeated infections, which are often not the case for virus-triggered diseases.

The above problems could be resolved by assembling a network of the transportation system on top of the local evolution of epidemics and formulate open-system compartments, which will bring the model resolution to the next stage. On this network, nodes represent population districts (communities/cities/countries) and edges describe transportation availability between nodes. An identical compartment model (e.g., SEIR model) is initiated at each node, which generates its own dynamics of the epidemic under the in- and out-flow of population that it exchanges from the overlying transportation network. Among various methodologies in epidemic research \citep{R2007}, this modeling approach has been widely taken by previous studies, and simulators are built to study the spread of epidemics as well as help design corresponding control policies after the burst of SARS \citep{Het2004}, H5N1 \citep{Get2006}, H1N1 \citep{BH2013}, on a national \citep{Fet2005, Fet2006, Bet2009} or global \citep{Het2004} scale. In these studies, the transportation system under concern is often considered as aggregated or single-layer \citep[e.g., the airline network,][]{Cet2006} with few exceptions \citep[e.g.,][]{Bet2009}, and people make simplifying assumptions to determine the flow matrix, which is often considered as stochastic \citep[e.g.,][]{Het2004}. 

Based on existing works, one notes that in a finer resolution, the public transportation system could be further modeled as a \textit{multi-layer} network \citep{Det2013,Bet2014,Det2016}, with each layer representing a specific means of transportation \citep{KT2006,Zet2010,Cet2013, GB2014,TP2015,CH2015, Aet2017}. On this network, nodes are maintained at different layers, in which case the network is sometimes termed as ``multiplex'' networks \citep{Net2013,NL2015,Set2016}. As in the single-layer representation, transportation takes place along network edges \citep{MB2012}, captured by a set of flowmaps that record the flow between each pair of connected nodes on each layer. One expects that the multi-layer representation of the transportation system outstands the aggregated single-layer representation since by differentiating means of transportation, the model could account for different diffusion properties in the spread of epidemics; from the perspective of policy analysis, such an increase of model resolution might provide valuable insights.

In this study, upon a multi-layer network model for the Chinese inter-city public transportation system, a simulator for the spread of epidemics in Chinese cities is built. On each of the over 340 prefectural-level cities, an identical SEIR model similar to the type in \citep{Wet2020} is assembled to characterize the local dynamics of the disease. Per the above discussion, the flow model on the network accounts for a set of important realistic concerns, including the transit of inbound flow, the cross-infection on public transportation media due to path overlap, the deactivation of outflow of the infected population, and the unlikelihood of repeated infections. Inspired by the real-world situation of Chinese administrative districts, the model also adopts a \textit{bi-partite} structure that partitions nodes into \textit{central cities} and \textit{peripheral cities}; this division determines the flow strength between connected nodes. Essentially, this multi-layer network model could serve as a general-purpose simulator for the evolution of epidemics in China (and potential other countries as well), which extends beyond the spread of 2019-nCov in Wuhan. Given the epidemiological parameters (basic reproduction number, incubation period, infection period, zoonotic force), geological sources (single location or multiple locations) and occurrence time (which determines the windowing of the seasonal flowmap) of a specific epidemic, the model could simulate the spread of the disease on all prefectural-level cities in China. With edge connectivities and the flowmaps being easily updated according to potential changes in public transportation infrastructures and in the distribution of population, this modeling framework could be of significant use for future policy decision-making on the control and emergency response for epidemics in China. 

\section*{Model Setup}

Consider the transportation network $\bm{G}$. At each node $i\in \bm{G}$, an SEIR model \citep[e.g.,][]{S2000,N2010} is assembled, where the population $P_i$ is divided into 4 compartments (Susceptible, Exposed, Infected, Recovered): $P_i = S_i + E_i + I_i + R_i$. Adopting the notations in \citep{Wet2020}, in the base model, the dynamics are governed by:

\begin{equation}
\left\{
\begin{aligned}
\dot{S}_i & = - \frac{S_i}{P_i}(\frac{R_0}{D_I}I_i + z) \\
\dot{E}_i  & = \frac{S_i}{P_i}(\frac{R_0}{D_I}I_i + z) - \frac{E_i}{D_E}  \\
\dot{I}_i  & = \frac{E_i}{D_E} - \frac{I_i}{D_I} \\
\dot{R}_i & = \frac{I_i}{D_I},
\end{aligned}
\right.
\end{equation} 

where $R_0$, $D_E$, $D_I$ are the basic reproduction number, the incubation period and the infection period, respectively; $z =z(t)$ is the zoonotic force, which is only nonzero for a certain period at the source node of the epidemics. Note that the venue from $E_i$ directly leading to $R_i$ is cut off, which exists in the general SEIR model.

These dynamics work for a closed population which could not account for the spread of diseases due to the population flow between cities (nodes). The flow is maintained on the overlying transportation network $\bm{G}$ which is cast as multi-layer, with each layer representing a specific means of transportation, including airline (A), railway (R), sail route (S) and bus (B). We denote the set of layers as $Q = \{A,R,S,B\}$. These layers share the same set of nodes $V=\{i\}$, the 347 prefectural-level cities in China, while having different sets of edges $L_q$ for $q\in Q$ according to the availability of transportation infrastructures. Edges are undirected since the transportation between two cities is bi-way. Note that we do not have a separate layer for the (private) car travel and instead assume that cars share the routes in the bus layer (see following sessions for further discussion). International population flow are not included in the current model, which are mainly carried out by airlines. Therefore we have $\bm{G} = \{G_A,G_R,G_S,G_B\}$, and:
\begin{equation}
\bm{G} =\left\{
\begin{aligned}
G_A & = (V,L_A) \\
G_R & = (V,L_R) \\
G_S & = (V,L_S) \\
G_B & = (V,L_B) 
\end{aligned}
\right.
\end{equation} 

Two properties differentiate means of transportation, besides layers' specific edge connectivities determined by the availability of infrastructures. First, for each $q$, there is a specific transfer rate $TR_i^q\in [0,1]$ for every city $i$, which represents the proportion of inflow to city $i$ that will be in transit and leave for the next destination, thus not entering the population stock $P_i$. Such transfer rates will be small for small cities and large for cities that are regional transportation centers. Ideally, the transfer rate will also be higher for certain transportation means such as railways and lower for others, say airlines (direct flights are primary among domestic air travels); yet for simplicity in the current model we assume that it is the same for all means of transportation at a certain city, i.e., $TR_i^q=TR_i$. Second, on different transportation media the likelihood that an exposed person will infect others during travel is not the same, which is represented by an \textit{in-travel} basic production number $R_T^q>1$. Expectedly, on airplanes, such a number will be lower than that on trains or buses where people talk more often, and in cars, the number is essentially 0; it is safe to assume that $R_T^q$ is homogeneous on different edges of layer $q$, i.e., the value only depends on the means of transportation. Features of the multi-layer network are illustrated in Figure 1.

\begin{figure}
\centering   
	\includegraphics[width=3.5in]{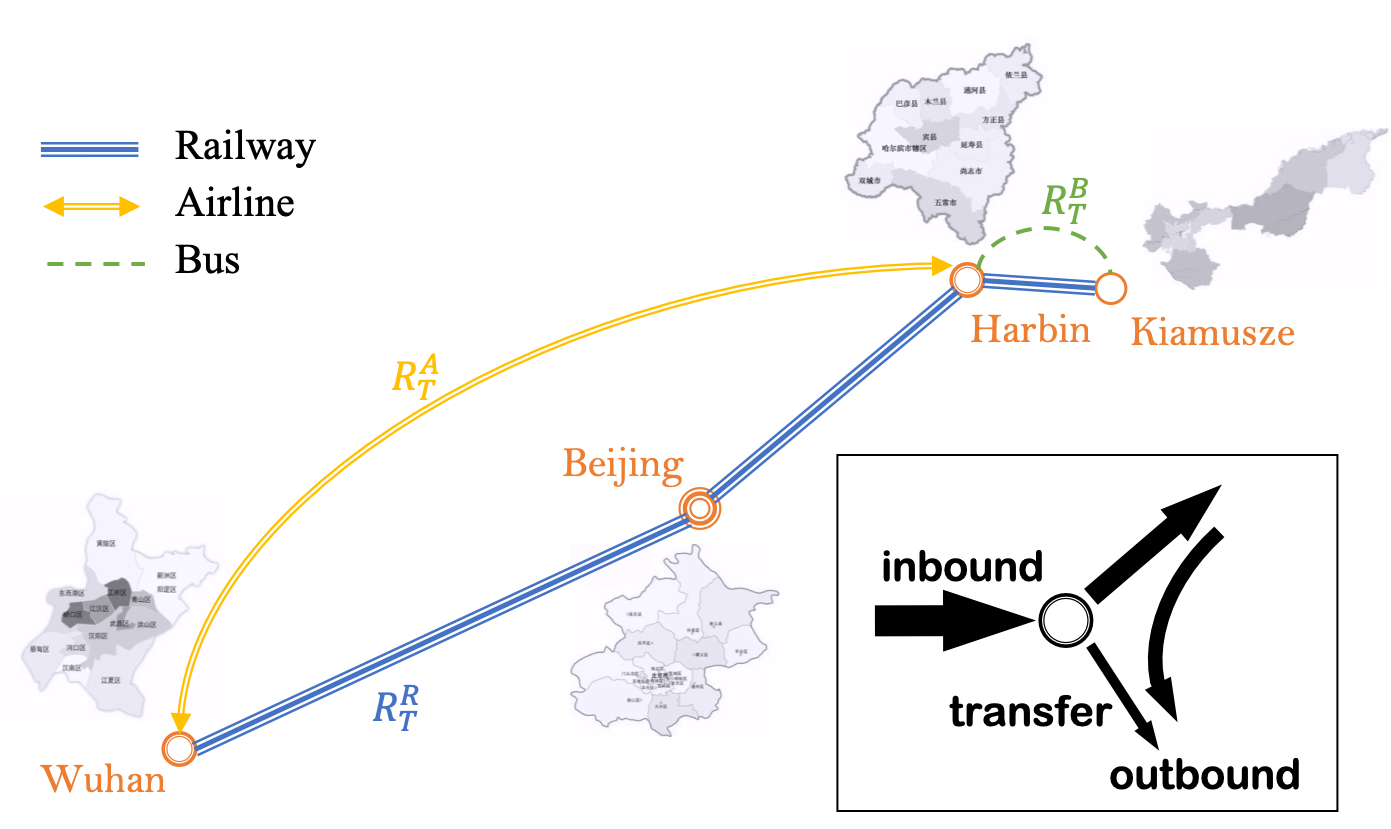}  
\caption{Illustration of the multi-layer inter-city transportation network. People move between cities (nodes) by different means of transportation, possibly with multiple hops on the network across different layers. Cross-infection takes place during the travel, with different transportation media having different levels of transmissivity $R_T$. Each city $i$ is associated with a specific transfer rate $TR_i$, the fraction of the inbound flow leaving for the next destination upon arrival.} 
\end{figure}

The in-travel transmissivity $R_T$ characterizes the cross-infection of the disease during public transportation. Importantly, cross-infection has spillovers due to path overlaps on public transport media, which deals with the topologies of network layers: a patient traveling from city $j$ to city $i$ by railway will likely spread the virus to everyone in the train during the travel between $j$ and $i$, including those who might get on the train at a different city $k$. This effect is pervasive in the real world, especially in China where the public transportation is crowded, and could be exempted only for an end-to-end transportation (e.g., airlines, cars). To model this path-dependent feature, we make two assumptions here (Figure 2). First, it is assumes that only those people that get off the train at the same destination $i$ as the patient will be infected. Implicitly, this is assuming that passengers sharing the same destination are more likely to stay close during the travel and the getting-off, which does not deviate from realistic situations. In a complete manner, the spillover will likely take place between any two routes that share a finite part; yet this will open a new dimension in the computation of the spillover and incur redundant complexity. Second, we assume that the strength of the spillover effect is proportional to the ratio of the (shortest path) distance between city $k$ and $i$ and the distance between $j$ and $i$, for a node $k$ that lies between $j$ and $i$ in the path; for a node $k$ that lies beyond $j$ in the path (i.e., earlier in the path), the $k\rightarrow i$ distance is thresholded by the $j\rightarrow i$ distance.

\begin{figure*}
\begin{minipage}{\linewidth}
\centering   
	\includegraphics[width=5in]{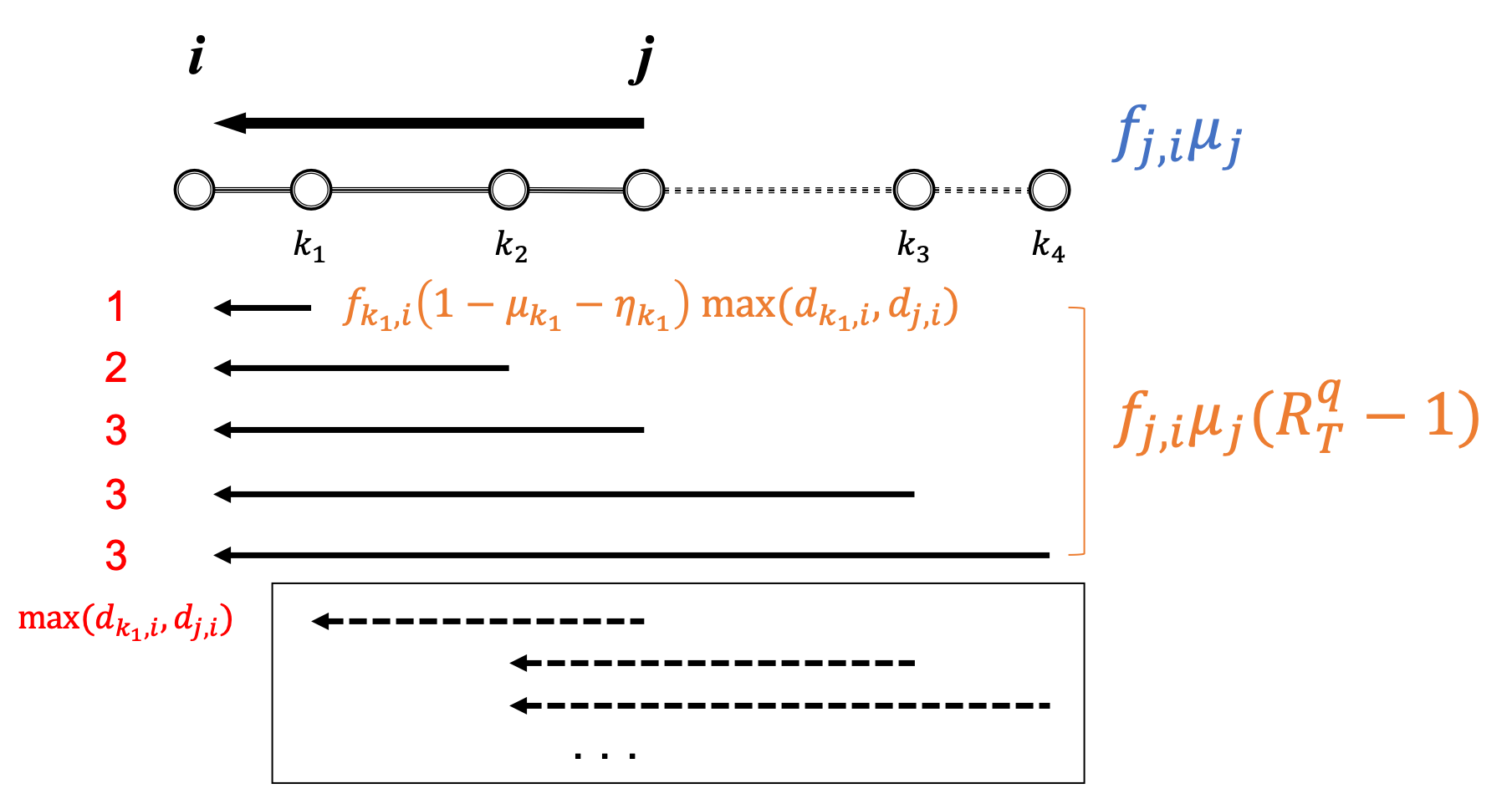}  
\caption{Illustration of the spillover of cross-infection during travel due to path overlap. If there are exposed cases in the population flow from city $j$ to city $i$, then cross-infection takes place on all routes $k\longrightarrow i$ that shares a part with the route $j\longrightarrow i$. The spillover $f_{j,i}^q \mu_j (R_T^q -1)$ of exposed cases is allocated proportionally to the susceptible population $f_{k,i}(1-\mu_k-\eta_k)$ among the outflow of city $k$, weighted by the shared length of the two route. Routes that share a finite part with $j\longrightarrow i$ but not terminate at $i$ are not considered in the spillover.} 
\end{minipage}
\end{figure*}

Base on the established model setting, we formulate the flow equations. On the transportation network, we track the in- and out-flow of the exposed (E), susceptible (S) and recovered (R) population of each city. For each means of transportation $q$, the flowmap between nodes is characterized by a matrix $F^q = \{f_{j,i}^q\}$; the matrix is not guaranteed to be symmetric since the flow $f_{i,j}^q$ from city $i$ to $j$ is not equal to the flow $f_{j,i}^q$ from city $j$ to $i$. Using values on the $|Q|=4$ flowmaps, at time $t$, the inflow of the exposed population to city $i$ summarize over all cities on the transportation network and over all means of transportation:
\begin{equation}
\Delta E_i^{in}(t) = \underset{q\in{Q}}{\sum} \underset{j\in{V}}{\sum} \overline{f_{j,i}^q (t) \mu_j (t)} (1-TR_i),
\end{equation}  
with the fraction of people in transit deducted. Here
\begin{widetext}
\begin{equation}
\overline{f_{j,i}^q  \mu_j}= f_{j,i}^q  \mu_j + \underset{k}{\overset{p^q(i,k)\cap p^q(i,j)\neq 0}{\sum}} f_{k,i}^q \mu_k(R_T^q -1)\frac{f_{j,i}^q(1-\mu_j-\eta_j)\text{min}(d^q_{j,i},d^q_{k,i})}{\underset{l}{\overset{p^q(i,k)\cap p^q(i,l)\neq 0}{\sum}} f_{l,i}^q (1-\mu_l-\eta_l)\text{min}(d^q_{l,i},d^q_{k,i})}
\end{equation}
\end{widetext}
is the adjusted exposed population flow from city $j$ to $i$ by means $q$, taking care of the spillover effect in cross-infection, in which $d^q_{i,j}$ represents the shortest path distance between $i$ and $j$ on layer $q$. And we have
\begin{equation}
\label{eq:mu}
\mu_i(t) = \frac{\Delta E_i^{out}(t) + \underset{q\in{Q}}{\sum} \underset{j\in{V}}{\sum} \overline{f_{j,i}^q (t-1) \mu_j (t)} TR_i^q}{\underset{q\in{Q}}{\sum} \underset{j\in{V}}{\sum} f_{i,j}^q (t)},
\end{equation}

and 
\begin{equation}
\label{eq:eta}
\eta_i(t) = \frac{\Delta R_i^{out}(t) + \underset{q\in{Q}}{\sum} \underset{j\in{V}}{\sum} f_{j,i}^q (t) \eta_j(t-1) TR_i^q}{\underset{q\in{Q}}{\sum} \underset{j\in{V}}{\sum} f_{i,j}^q (t)},
\end{equation}
which are the time-stamped proportion of the exposed and recovered population among the total outflow population of city $i$, respectively. Therefore, $(1-\mu_i-\eta_i)$ is the proportion of the susceptible population among the total outflow of city $i$. The exposed population among the total outbound of $i$ is contributed by both the exposed originated from city $i$ ($\Delta E_i^{out}$, see below) and the exposed among the transit of inbound (second term in the numerator of $\mu_i$, proportion to $\Delta E_i^{in}$ by a term of the transfer rate), and is dynamically updated at each step; we denote the proportion by $\mu_i(t)$. Similarly, we track the stock (via proportion $\eta_i(t)$) of the recovered population among the total outflow of a city, which is also updated at each step. This tracking is an important caveat in this model since we consider that a recovered person will unlikely to be infected again. Therefore, by this means, only the susceptible population $(1-\mu_i-\eta_i)$ among the total outflow along each edge will be subject to cross-infection during travel.

The outflow population from city $i$'s population $P_i$ is the total outbound flow minus the transferred inbound flow, which is contributed to by the $S,E,R$ compartments (not $I$), different from the case in \citet{Wet2020}, where it assumes that all four compartments contribute to the outbound flow of Wuhan. Proportionally, the outflow of the exposed is:
\begin{equation}
\Delta E_i^{out}(t) = E_i(t) \frac{ \underset{q\in{Q}}{\sum} \underset{j\in{V}}{\sum} f_{i,j}^q (t) - \underset{q\in{Q}}{\sum} \underset{j\in{V}}{\sum} f_{j,i}^q (t) TR_i^q}{S_i(t) + E_i(t) + R_i(t)}.
\end{equation}  

Note that the flowmaps and the transfer rates should ensure that at each city, the inbound transfer flow should always be smaller than the outbound flow, i.e.,:
\begin{equation}
\label{eq:ineq}
\begin{aligned}
&\underset{q\in{Q}}{\sum} \underset{j\in{V}}{\sum} f_{i,j}^q (t) - \underset{q\in{Q}}{\sum} \underset{j\in{V}}{\sum} f_{j,i}^q (t) TR_i^q > 0 \\
&\Longrightarrow TR_i^q < \frac{\underset{q\in{Q}}{\sum} \underset{j\in{V}}{\sum} f_{i,j}^q (t)}{ \underset{q\in{Q}}{\sum} \underset{j\in{V}}{\sum} f_{j,i}^q (t)},\ \ \forall\ i\in V, q\in Q.
\end{aligned}
\end{equation} 

For the recovered population, the inflow is tracking all the recovered people (through $\eta_j$) upon arrival:
\begin{equation}
\Delta R_i^{in}(t) = \underset{q\in{Q}}{\sum} \underset{j\in{V}}{\sum} f_{j,i}^q (t) (1-TR_i^q) \eta_j(t-1),
\end{equation}  
and the outflow is proportional to $\Delta E_i^{out}$:
\begin{equation}
\Delta R_i^{out}(t) = R_i(t) \frac{ \underset{q\in{Q}}{\sum} \underset{j\in{V}}{\sum} f_{i,j}^q (t) - \underset{q\in{Q}}{\sum} \underset{j\in{V}}{\sum} f_{j,i}^q (t) TR_i^q}{S_i(t) + E_i(t) + R_i(t)}.
\end{equation}  

For the susceptible population, according to the flow balance, $\Delta S_i^{in}$ is the total un-transferred inflow subtracting the recovered and the exposed inflow:
\begin{equation}
\Delta S_i^{in}(t) = \underset{q\in{Q}}{\sum} \underset{j\in{V}}{\sum} f_{j,i}^q (t) (1-TR_i^q) - \Delta E_i^{in}(t) - \Delta R_i^{in}(t),
\end{equation}  
and
\begin{equation}
\Delta S_i^{out}(t) = S_i(t) \frac{ \underset{q\in{Q}}{\sum} \underset{j\in{V}}{\sum} f_{i,j}^q (t) - \underset{q\in{Q}}{\sum} \underset{j\in{V}}{\sum} f_{j,i}^q (t) TR_i^q}{S_i(t) + E_i(t) + R_i(t)}.
\end{equation}  

In the end, we arrive at the open-system SEIR model accounting for the transportation flow:
\begin{equation}
\left\{
\begin{aligned}
\dot{S}_i & = - \frac{S_i}{P_i}(\frac{R_0}{D_I}I_i + z)+ \Delta S_i^{in}  - \Delta S_i^{out}\\
\dot{E}_i  & = \frac{S_i}{P_i}(\frac{R_0}{D_I}I_i + z) - \frac{E_i}{D_E}  + \Delta E_i^{in}  - \Delta E_i^{out}\\
\dot{I}_i  & = \frac{E_i}{D_E} - \frac{I_i}{D_I} \\
\dot{R}_i & = \frac{I_i}{D_I} + \Delta R_i^{in} - \Delta R_i^{out}
\end{aligned}
\right.
\end{equation} 

with $\mu_i$ and $\eta_i$ updated according to \eqref{eq:mu} and \eqref{eq:eta}. Note that in the model we don't assume a balance of the inbound and outbound flow at each city; there is no such equilibrium for the population flow in real practice, and each city's transient population is in constant dynamics. However, the inequality \eqref{eq:ineq} regarding the inbound transit and the outbound should always hold at each city. 

\section*{Reduction of the Parameter Space}
The model incorporates 4 layers of the public transportation system (airline, train, sail, bus), which corresponds to the situation in China but may be applicable to other countries as well. Based on empirical observations, simplifying assumptions on the topologies of the transportation network in China are to be made. Most importantly, the multi-layer network could be further represented by a bi-partite structure (i.e., factor graph \citep{Zet2019}), with nodes $V$ divided into two categories: \textit{central cities} $V_c$ and \textit{peripheral cities} $V_p$ (Figure 3). Upon this division, we assume different flow strength on edges between $V_c\leftrightarrow V_p$, $V_c\leftrightarrow V_c$ and $V_p\leftrightarrow V_p$ when constructing the flowmaps $F$. We are also able to assign a homogeneous value of the transfer rate $TR$ to each category of nodes. These treatments bring down the model's parameter space to a feasible region (2 for $TR$, 4$\times$3 = 12 for $F$). This division of cities' roles is valid in China, since Chinese cities are often categorized on a level base \citep{Note1}. In the current study we apply a two-bin partition and regard all provincial capital cities and level 0-2 cities as central cities $V_c$ (54 counts, Table 1), with the rest 293 cities belong to $V_p$; expectedly, in a finer scale, cities could be categorized into more bins, and the flowmap will then have a multi-partite structure similar to the celebrated stochastic block model \citep{Het1983}.

\begin{figure*}
\begin{minipage}{\linewidth}
\centering   
	\includegraphics[width=5in]{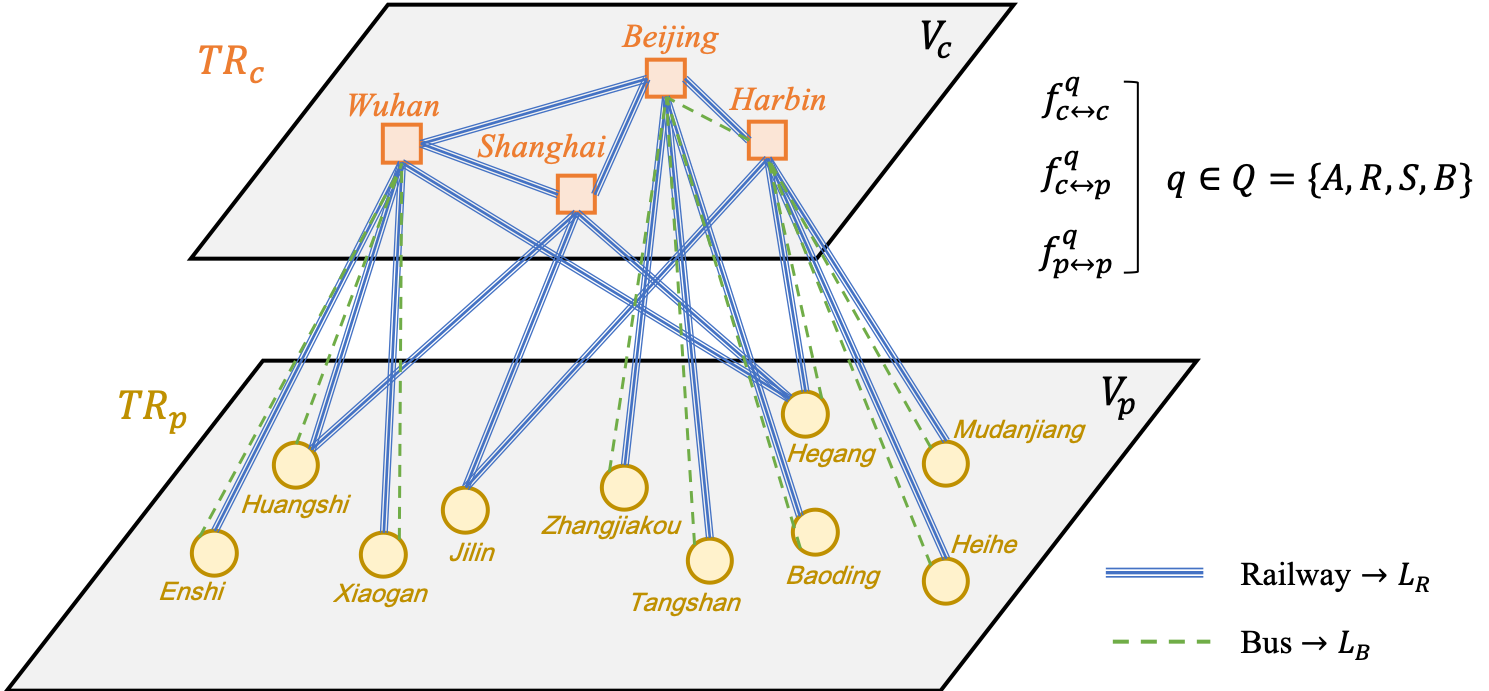}  
\caption{Illustration of the multi-layer bi-partite structure of the transportation network. Cities (nodes) are categorized as \textit{central cities} $V_c$ and \textit{peripheral cities} $V_p$. For demonstration, only two layers $L_R$ and $L_B$ are shown. Bi-partite structure facilitates the determination of the transfer rate $TR$ and the flowmaps $F$.} 

\end{minipage}
\end{figure*}

When real datasets could be acquired, these parameters may be determined in a more objective manner. Ideally, the flow $f_{j,i}^q$ from node $j$ to node $i$ by means $q$, is supposed to be cast as a time series that incorporates seasonal features but approximately remains invariant on the annual basis. Then for a specific starting time of an epidemic, a cursor is placed on the annual curve and a window of the flowmap is then cut out from this point on for the usage in simulation. Moreover, in cases where only the aggregated flow of all transportation layers between two cities $F_{i,j}= \sum_q \{f_{j,i}^q\}$ is available, one may apply a multinomial logit model to determine the flow for each layer, as in \citep{Zet2014}. For the transfer rate, if the city's (average/annual) transient population $P_i^T$ is available, then $TR$ could be calculated by comparing the aggregated outflow and the transient population:
\begin{equation}
TR_i = \frac{\underset{q\in{Q}}{\sum} \underset{j\in{V}}{\sum} f_{i,j}^q (t) - P_i^T}{\underset{q\in{Q}}{\sum} \underset{j\in{V}}{\sum} f_{j,i}^q (t)}.
\end{equation}
However, this detailed formulation might be subject to error when data quality is not guaranteed, since one has to ensure that the outbound is always greater than the transient population. 

\begin{figure*}
\begin{minipage}{\linewidth}
\begin{table}[H]
\resizebox{\textwidth}{!}{ %
\begin{tabular}{@{}|cccc||cccc||cccc||@{}}
\toprule
\textbf{City}&\textbf{$A_i$}&\textbf{Rank($B_i$)}&\textbf{Rank($C_i$)}&\textbf{City}&\textbf{$A_i$}&\textbf{Rank($B_i$)}&\textbf{Rank($C_i$)}&\textbf{City}&\textbf{$A_i$}&\textbf{Rank($B_i$)}&\textbf{Rank($C_i$)}\\ \midrule
\hline
\textbf{Shanghai*}   & 4   & 8 & 347  & \textbf{Zhengzhou*}   & 3   & 9 & 110   & \textbf{Yangzhou}   & 2  & 131 & 209     \\ \midrule

\textbf{Beijing*}   & 3   & 4 & 337  & \textbf{Yantai}   & 3   & 82 & 99   & \textbf{Taizhou}   & 2  & 181 & 88  \\ \midrule

\textbf{Guangzhou*}   & 4   & 1 & 318  & \textbf{Dongguan}   & 2   & 254 & 293   & \textbf{Jiaxing}   & 2  & 266 & 181  \\ \midrule

\textbf{Shenzhen}   & 4   & 20 & 291  & \textbf{Jinan*}   & 2   & 18 & 77   & \textbf{Xiamen}   & 3  & 26 & 8  \\ \midrule

\textbf{Tianjin*}   & 4   & 32 & 277  & \textbf{Quanzhou}   & 2   & 112 & 319   & \textbf{Jinhua}   & 3  & 63 & 101  \\ \midrule

\textbf{Suzhou}   & 3   & 233 & 338  & \textbf{Harbin*}   & 3   & 14 & 63   & \textbf{Nanning*}   & 4  & 17 & 29  \\ \midrule

\textbf{Chongqing*}   & 4   & 15 & 256  & \textbf{Nantong}   & 3   & 83 & 100   & \textbf{Huhhot*}   & 3  & 21 & 25    \\ \midrule

\textbf{Chengdu*}   & 3   & 3 & 237  & \textbf{Changchun*}   & 3   & 16 & 69   & \textbf{Zhongshan}   & 2  & 292 & 279  \\ \midrule
\textbf{Wuhan*}   & 4   & 5 & 222  & \textbf{Shijiazhuang*}   & 3   & 34 & 78   & \textbf{Huizhou}   & 1  & 270 & 258  \\ \midrule
\textbf{Hangzhou*}   & 3   & 28 & 176  & \textbf{Xi'an*}   & 3  & 6 & 39   & \textbf{Taiyuan*}   & 2  & 22 & 19  \\ \midrule
\textbf{Wuxi}   & 3   & 57 & 155 & \textbf{Fuzhou*}   & 3   & 33 & 50   & \textbf{Urumqi*}   & 3  & 2 & 23  \\ \midrule
\textbf{Qingdao}   & 3   & 40 & 177  & \textbf{Hefei*}   & 2   & 19 & 70   & \textbf{Yinchuan*}   & 3  & 29 & 26  \\ \midrule
\textbf{Nanjing*}   & 4   & 24 & 140  & \textbf{Xuzhou}   & 3   & 35 & 86   & \textbf{Guiyang*}   & 3  & 27 & 24  \\ \midrule
\textbf{Dalian}   & 3   & 31 & 123  & \textbf{Changzhou}   & 3   & 60 & 79   & \textbf{Lanzhou*}   & 3  & 13 & 30  \\ \midrule
\textbf{Foshan}   & 3   & 146 & 278  & \textbf{Wenzhou}   & 3   & 48 & 46   & \textbf{Haikou*}   & 3  & 30 & 4  \\ \midrule
\textbf{Shenyang*}   & 3   & 11 & 124  & \textbf{Shaoxing}   & 1  & 229 & 207   & \textbf{Zhuhai}   & 3  & 47 & 33  \\ \midrule
\textbf{Ningbo}   & 3   & 77 & 109  & \textbf{Kunming*}   & 3   & 7 & 22  & \textbf{Xining*}   & 2  & 12 & 44  \\ \midrule
\textbf{Changsha*}   & 3   & 10 & 98  & \textbf{Nanchang*}   & 3   & 23 & 51  & \textbf{Lhasa*}   & 2 & 25 & 174  \\ \bottomrule

\end{tabular}}%
\caption{Centrality measures of central cities $V_c$ in the transportation network, including all provincial capitals (stars) and level 0-2 cities in China.}
\end{table}
\end{minipage}
\end{figure*}

\section*{Connectivities of the Chinese Public Transportation System}

As in many countries, different layers of the public transportation system in China play different roles in serving the demand of population flow, and therefore edge connectivities are different for each layer \citep{Note2}. For airlines, we obtain public data on the airline schedules for 90 cities; for railways, there are 11 primary railroads in China which occupy the major railway passenger flow, connecting $\sim 120$ cities into the network. For sails, there are roughly 19 passenger lines along the east coast of China and the Yangtze river; yet sails take a very small part ($<2\%$) in Chinese public transportation. For buses, schedules are difficult to collect, but according to empirical observations, we construct edges on the bus layer between cities within a mid-range geological distance (150km), and between a provincial capital city to other cities in the province. The four layers ($G_A,G_R,G_S,G_B$) of the Chinese inter-city public transportation network are shown in Figure 4.

\begin{figure*}
\begin{minipage}{\linewidth}
\centering   
	\includegraphics[width=5.8in]{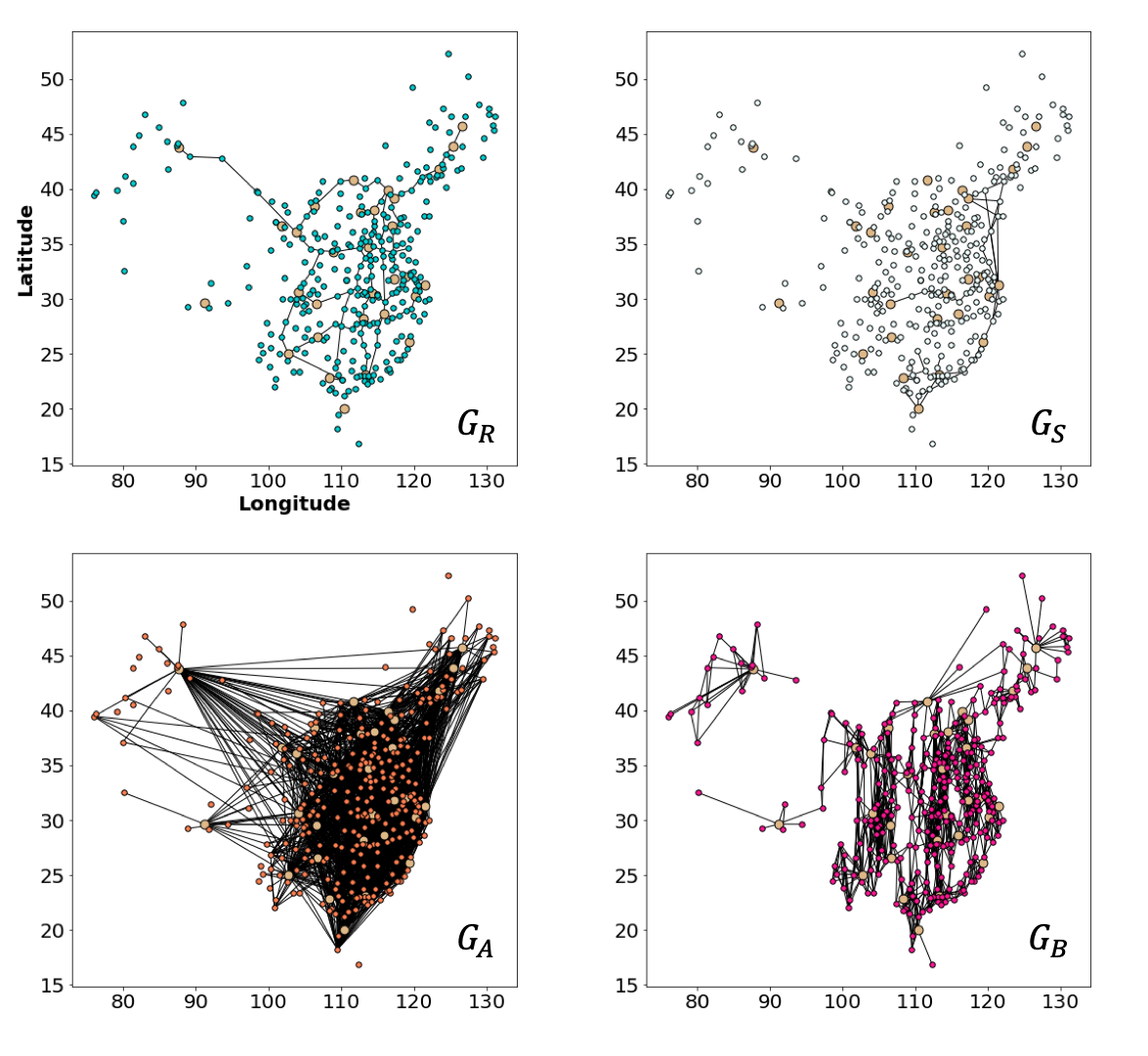}  
\caption{Edge connectivities of layers $G_A,G_R,G_S,G_B$, representing the multi-layer inter-city public transportation system in China. Big yellow nodes denote \textit{central cities} $V_c$; small nodes denote \textit{peripheral cities} $V_p$.} 
\end{minipage}
\end{figure*}

Based on the determined edge connectivities, we calculate nodes' centrality measures on the network, which could be used to distinguish central cities from peripheral cities. Three measures are computed for each node $i\in V$: node activity $A_i$ \citep{NL2015}, which is the count of layers in which node $i$ has at least one edge; betweenness $B_i$:
\begin{equation}
B_i = \underset{s\neq t\neq i}{\underset{s,t \in V}{\sum}} \frac{\sigma_{s,t}^{i}}{\sigma_{s,t}}.
\end{equation}
and coupling strength $C_i$ \citep{MB2012}, defined as:
\begin{equation}
C_i =  \underset{j \neq i}{\underset{j \in V}{\sum}} \frac{\sigma_{i,j}^{coupled}}{\sigma_{i,j}},
\end{equation}
where $\sigma_{s,t}$ in both cases indicates the number of shortest paths between nodes $s$ and $t$, $\sigma_{s,t}^{i}$ are the number of such shortest paths that traverse node $i$, and $\sigma_{s,t}^{coupled}$ is the number of shortest paths that use edges on more than one layer. These three measures characterize the centrality of nodes on a multilayer network from different perspectives: $A_i$ indicates the presence of $i$ on different layers, $B_i$ indicates its significance in bridging network paths, and $C_i$ indicates the extent of its embedding into the multiplexity of the network. Note that $B_i$ is calculated on the aggregated network that is the superposition of multiple layers. Results of these centrality measures for the 54 central cities are shown in Table 1, with their average values compared to those of peripheral cities (Figure 5). Many central cities are active on 3 or 4 layers, with an average $A_i$ of 2.87, compared with 1.65 for peripheral cities. The average betweenness $B_i$ of central cities are 35 times larger than that of peripheral cities, and ranking $\# 1-35$ largest betweenness are all on central cities. For coupling strength $C_i$, central cities and peripheral cities are almost indistinguishable. This is consistent with expectation, since on one hand, central cities have multiple means of transportation that likely increases its coupling strength, while on the other hand, central cities often have well-developed airlines that are always the shortest paths, which reduces $C_i$; these two competing forces decide that the coupling strength of central cities is not going to be clearly larger than that of peripheral cities. Overall, these results suggest that our hard division of $V_c$ and $V_p$ according to public information is reasonable on the basis of real edge connectivities.

\begin{figure}[!htp]
\centering   
	\includegraphics[width=3.5in]{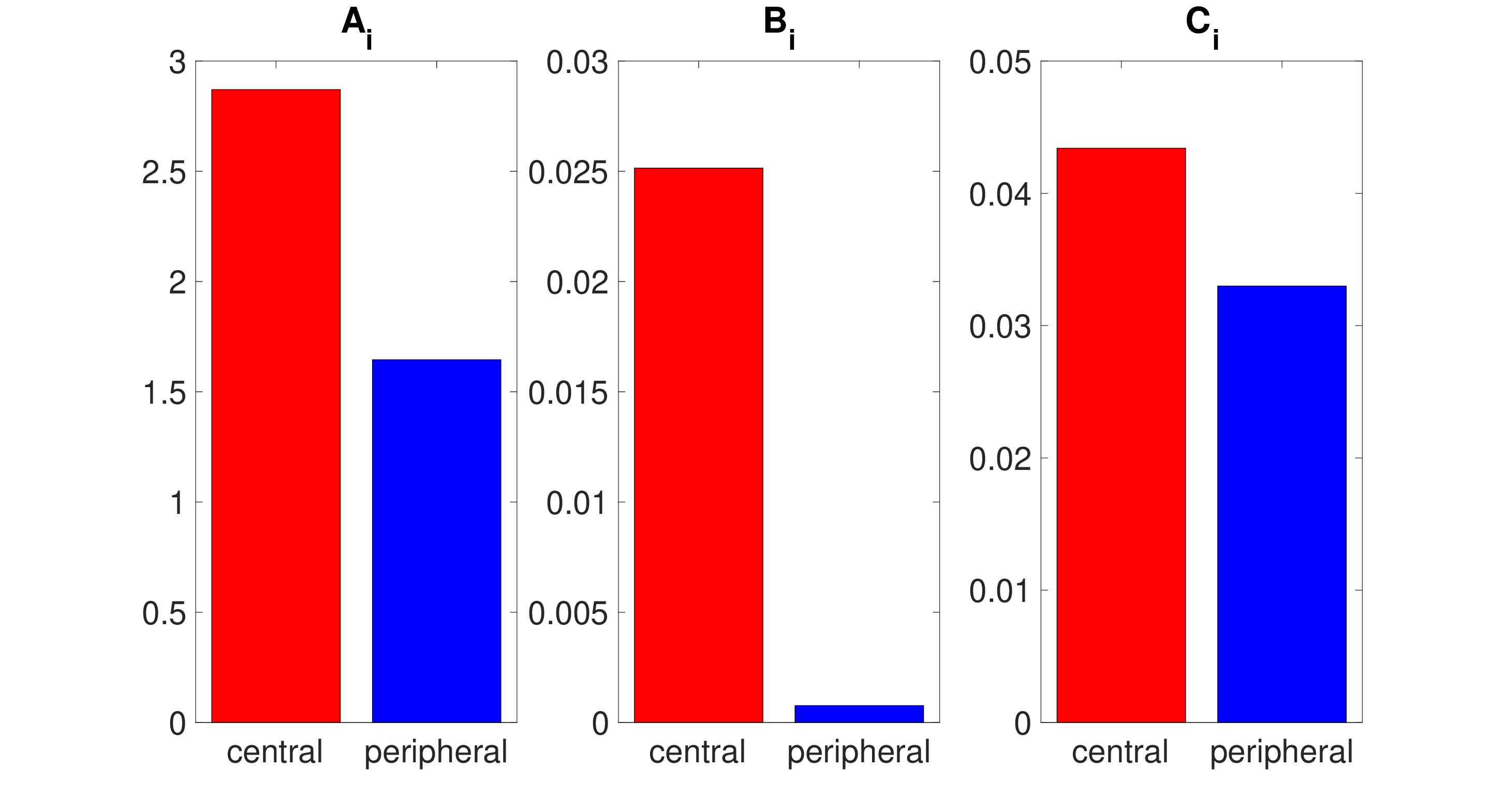}  
\caption{Group average of centrality measures for central cities (red) and peripheral cities (blue). Central cities have greater node activity $A_i$ and betweenness $B_i$ than peripheral cities; coupling strength $C_i$ is similar for two categories of nodes.} 
\end{figure}

\section*{Simulating the Wuhan 2019-nCov Epidemic}

We use the model to simulate the spread of the Wuhan Coronavirus in Chinese cities under assumed parameters determined from empirical considerations. For epidemiological parameters, we let $R_0 = 2.68$, $D_E = 6$ days, $D_I = 2.4$ days, according to \citet{Wet2020}; we starts the simulation at Dec. 8th, 2019, when the first case was reported, with a nonzero zoonotic force $z = 27$ cases/day (the total number of confirmed cases in December) before Jan. 1st the date of the closure of the Wuhan Huanan seafood market, which is the claimed source of the virus. We tested the values suggested by \citet{Wet2020} ($z = 86$ cases/day extending 30 days) and results show a misfit (see below) three times larger than using our values for $z$; therefore the values in \citet{Wet2020} are not used. For parameters of the transportation network, the in-travel basic production number $R_T$ is assumed to be: $\{R_T^A,R_T^R,R_T^S,R_T^B\} = \{1.2,1.5,1.5, 3\}$, smallest for airlines, largest for buses. Flowmaps are determined based on empirical constraints (all units are people/day). Airlines are the major transportation means between central cities, and we determine $f_{cc/cp/pp}^A = 1000/500/0$, assuming no airline transport between peripheral cities, four flights between central cities per day, and half the amount between a central city and a peripheral city. For railways the flow is determined as $\hat{f}_{cc/cp/pp}^R = 2000/200/500$, and $f_{i,j} = \hat{f}_{i,j}/d_{i,j}$, i.e., passenger flows are inversely proportional to the shortest path distance between cities, under the carriage capacity 2000 people/train. For sail routes, $f_{cc/cp/pp}^S = 100/100/100$, which occupy a small fraction of the total flow. For buses, we let $f_{cc/cp/pp}^B = 0/3000/1000$, assuming no bus travel between central cities; $f_{pp}^A = 0$ and $f_{cc}^B=0$ reinforce the bi-partite structure of the model. Buses are heavily used in provincial transport, for travelers in peripheral cities to go to either the local central city or nearby peripheral cities. We let the numbers (3000 and 1000) be three/two times the normal flow (50 people/bus, 20 or 10 buses between central-peripheral and peripheral-peripheral) to account for the private transportation by cars that nevertheless contributes to the entire population flow, which is mostly provincial rather than regional (to a nearby province). Implicitly, as car travels are aggregated into bus travels, the $R_T^B$ represents the effective value watered down from a real transmissivity on buses. Note that the assumed constant flowmaps do not consider seasonal effects, which might be significant in certain cases, exactly like for the Wuhan epidemic which took place right before the Chinese New Year when massive transports are carried out. Although not obtained from a full-parameter inversion, the values for $R_T$ and $F$ that we use are nevertheless well tested by extensive forward simulation runs and are confirmed as quasi-locally-optimal values.

The current runtime of the simulator is approximately 55 seconds per simulation time step, estimated on a 2018 Macbook Pro; for a simulation range of 50 time steps, one forward run takes rough 0.75 hours. Given the large parameter space even for the simplified model (2 for $TR$, 12 for $F$, 4 for $R_T$ and 5 epidemiological parameters), a full-parameter inversion is not feasible on personal computing devices. Per the above discussion, with other parameters determined through empirical considerations, $TR_c$ and $TR_p$ are set as open parameters and we initiate a partial inversion.  At each run, the model is simulated for 53 time steps (days), from the date of the first case (Dec. 8) minus an incubation period (6.4 days), to the day before the Chinese New Year's Eve (Jan. 23). This stopping time is reasonable since most population flow in China during winter travels took place before this date, right after which the government took urgent measures and asked the entire domestic population to self-quarantine and abandon inter-city public transportation. After Jan. 23, confirmed cases of the epidemic are collected from each prefectural-level city and are reported to the public; we use this dataset for inversion \citep{Note3}. We observe that the fraction of confirmed cases in each city $\rho$ among national headcounts remains roughly invariant after around Feb.15th (Figure 6) due to the reduction of daily new headcounts as the effect of nationwide measures, with Wuhan having $\sim 60\%$ of all confirmed cases and all cities in the Hubei Province having $>80\%$ cases. Therefore, instead of matching the simulated time series with the entire data series, for the inversion we minimize the misfit between the simulated and realistic $\rho$ obtained at the end of the two time series, summing over each city except Wuhan in prevention of double-counting because $\sum_i\rho_i = 1$. This choice of misfit is also consistent with the fact that we don't have a reliable estimate of the zoonotic force, and also that the transportation network setting focuses more on the relative strength of the epidemic in each city. Since the simulation stops at Jan.23 but the confirmed cases are reported gradually afterwards, when calculating the $\rho$ of the simulation result we regard the exposed (E) as additional infected (I) cases as they will eventually enter the stock, a valid treatment given that the incubation period is unlikely to be as long as 30 days (from Jan. 23 to Feb. 23).

\begin{figure}
\centering   
	\includegraphics[width=3.5in]{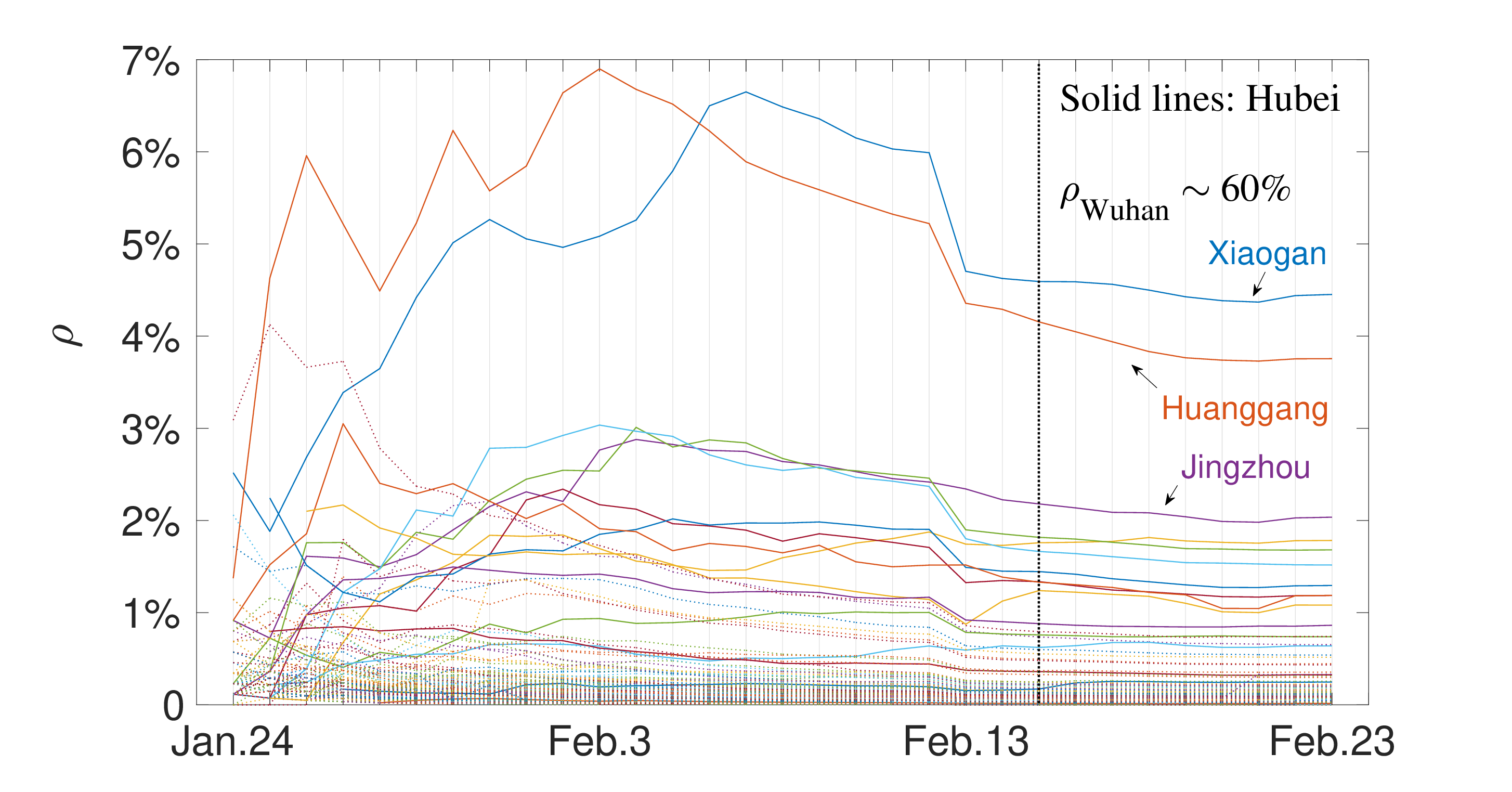}  
\caption{Fraction of the confirmed cases of 2019-nCov in each prefectural-level city among national headcounts as a function of time. Numbers become roughly invariant after around Feb.15th, suggesting a slow-down of the epidemic. Wuhan has $\sim 60\%$ of all cases; top three cities having the largest headcount besides Wuhan are Xiaogan, Huangshi and Jingzhou, all in the Hubei Province.} 
\end{figure}

Although real datasets for flowmaps are not applicable and a full-parameter inversion is not conducted, quite surprisingly, results of the coarse partial inversion nevertheless demonstrated great fitting performance of our model. The best-fit $TR_c = \textbf{14.95\%}$, $TR_p = \textbf{0.36\%}$, i.e., on average around 15$\%$ inbound flow at central cities are in transit, and the number is near 0 for peripheral cities, which is largely consistent with the empirical observation. We show the best-fit $\rho$ of Chinese provincial districts (except Hubei; Hong Kong, Macau and Taiwan not shown), aggregated from prefectural-level cities (left, Figure 7). It suggests that our results (red dots) recover the data (bars) to a satisfying extent: the severe situation of the disease in Hehan, Hunan, Shandong and Jiangxi is correctly revealed, while the spread in provinces such as Chongqing, Anhui, Zhengjiang, Guangdong and Heilongjiang is underestimated. This result might be consistent with the public news that after the burst of the epidemic there are uncommonly large population flow directed into these provinces from Wuhan, a situation not captured by our simplified flowmaps. Entire best-fit time series are also shown (right, Figure 7), on five example cities: Wuhan (origin of the epidemic), Beijing (capital of China), Huanggang (peripheral city in Hubei), Harbin (central city outside Wubei) and Kiamusze (peripheral city outside Wubei). The data (solid lines) are S-shape as the epidemic is gradually under control after Chinese New Year, while the simulation time series (dash lines) are demonstrating exponential growth, as one expects from the SEIR model. Nevertheless, applying a uniform time-shift ($\Delta t = 20$) assuming implicitly that all cities took measures at the same time, which was roughly the case in China, simulation results match the initial part of the real S-curve by a large margin. However, the matching of the two curves should not be over-interpreted as the model is always able to generate exponential growth and the value of the time-shift is not warranted. The absolute values of the confirmed cases, instead of the fraction $\rho$, are also fit reasonably well (bars in Figure 7 right), suggesting that our choice of model parameters makes certain sense.
Overall, given the coarse treatment during the parameter determination and data-processing process, these partial inversion results are believed to be acceptable, suggesting that the model is promising in generating reference dynamics for the spread of epidemics in China.

\begin{figure*}
\begin{minipage}{\linewidth}
\centering   
	\includegraphics[width=7.2in]{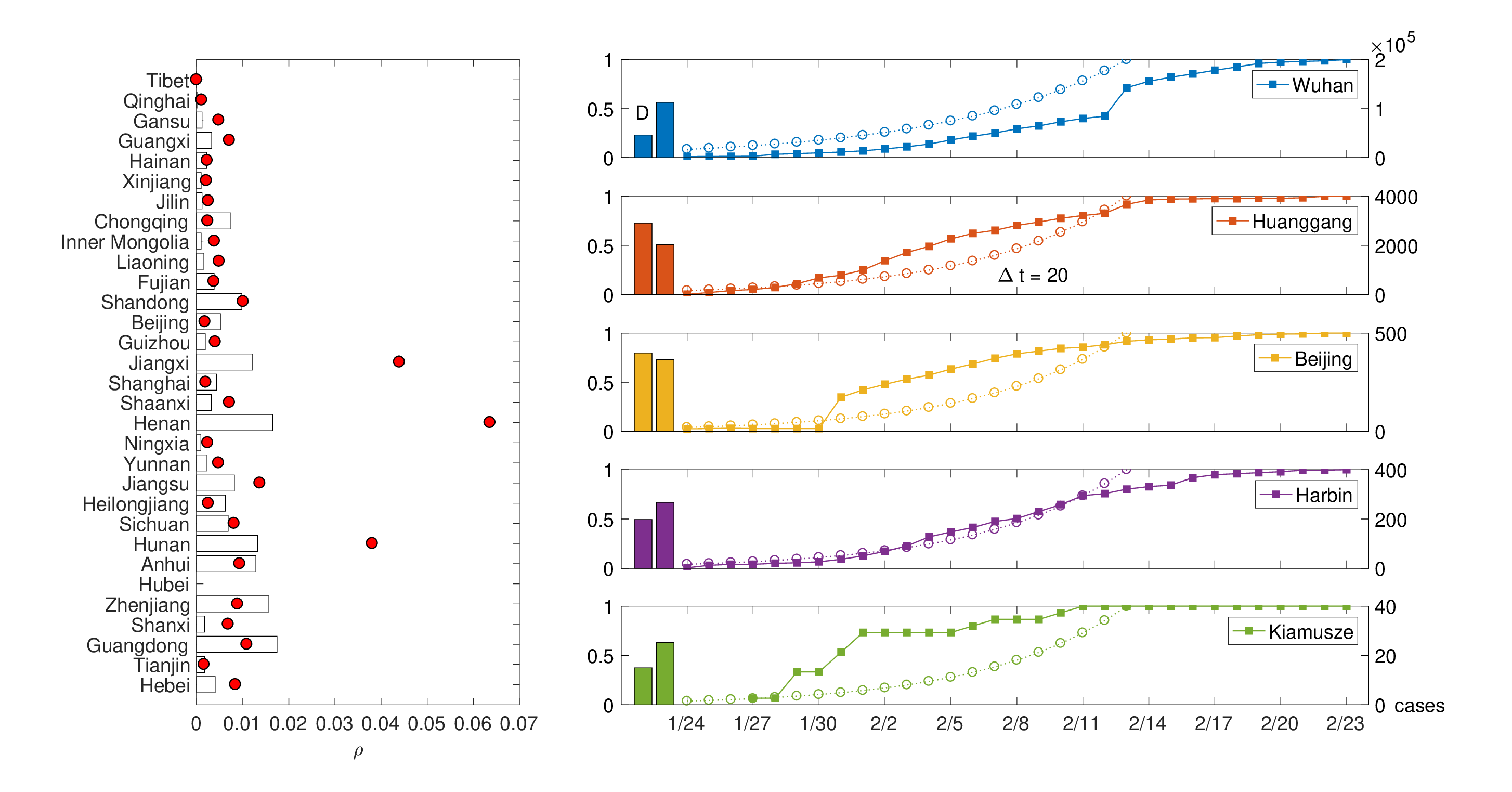}  
\caption{Inversion results. Left: fraction of confirmed cases in Chinese provincial districts except Hubei (Hong Kong, Macau and Taiwan not shown). Data: bars. Best-fit results: red dots. Right: normalized time series of the epidemic (left y-axis) and the absolute number of confirmed cases (bar, right y-axis) in Wuhan (origin), Beijing (capital), Huanggang (peripheral city in Hubei), Harbin (central city outside Wubei) and Kiamusze (peripheral city outside Wubei). Data: solid lines, first bar on the left. Best-fit results: dash lines, second bar on the left. Simulation time series are shift to the left by $\Delta t = 20$.} 
\end{minipage}
\end{figure*}

\section*{Concluding Remarks}

For this study, an SEIR model is used as the baseline epidemic model; therefore, as discussed earlier, this simulator is only reliable to generate dynamics for the period where no effective government intervention has been implemented. After actions are taken, the SEIR compartments will be invalidated, and more elaborated compartment models should be adopted to account for government's measures such as quarantine and the reportage of suspected cases \citep{Zet2005}, as well as the shutdown of transportation at certain places. Another extension to the current model framework is to relax the assumption in the spillover of cross-infection, and instead allow that cross-infection occurs between all routes sharing a finite part in the path. This will increase one search depth in the computation and will be feasible on massive-scale clusters; a super computing device will also facilitate a full-parameter inversion for the Wuhan coronavirus, ideally with real datasets assembled for flowmaps, whose results will undoubtedly uncover further information for the study of this on-going epidemic. Overall, constructed on the multi-layer network flow model with flexible inputs of edge connectivities, flowmaps and arbitrary system parameters, this general-purpose simulator for the city-level spread of epidemics in China has an adaptive nature that could be tuned for specific usage, which might be helpful for policy analysis in emergence response, and early-warnings of future events.

\end{document}